\documentstyle[preprint,aps,eqsecnum]{revtex}
 
\def\beq#1{\begin{equation} \label{#1}}
\def\eeq{\end{equation}}

\input prepictex
\input pictex
\input postpictex
\newdimen\tdim
\tdim=\unitlength
\def\stpltsmbl{\setplotsymbol ({\small .})}

\newbox\phru
\setbox\phru=\hbox{\beginpicture
\setcoordinatesystem units <\tdim,\tdim>
\stpltsmbl
\setquadratic
\plot
0 0
2.5 3
5 0
7.5 -3
10 0
/
\endpicture}
\def\photonru #1 #2 *#3 /{\multiput {\copy\phru}  at
#1 #2 *#3 10 0 /}
 
\newbox\sru
\setbox\sru=\hbox{\beginpicture
\setcoordinatesystem units <\tdim,\tdim>
\stpltsmbl
\setquadratic
\plot
  0.0   0.0
  4.8   1.5
  7.5   5.0
  7.3   8.5
  5.0  10.0
  2.7   8.5
  2.5   5.0
  5.2   1.5
 10.0   0.0
/
\endpicture}
\def\springru #1 #2 *#3 /{\multiput {\copy\sru}  at
#1 #2 *#3 10 0 /}

\begin{document}
{
\tighten
 
\title{Search for New Physics in  $D^\pm \rightarrow K_S X^{\pm} $ 
and $D^\pm \rightarrow K_S K_S K^{\pm}$ }
 
\author{Harry J. Lipkin\,\thanks{Supported
in part by The German-Israeli Foundation for Scientific Research and 
Development (GIF) and by the U.S. Department
of Energy, Division of High Energy Physics, Contract W-31-109-ENG-38.}}
\address{ \vbox{\vskip 0.truecm}
  Department of Particle Physics
  Weizmann Institute of Science, Rehovot 76100, Israel \\
\vbox{\vskip 0.truecm}
School of Physics and Astronomy,
Raymond and Beverly Sackler Faculty of Exact Sciences,
Tel Aviv University, Tel Aviv, Israel  \\
\vbox{\vskip 0.truecm}
High Energy Physics Division, Argonne National Laboratory,
Argonne, IL 60439-4815, USA\\
~\\HJL@axp1.hep.anl.gov
\\~\\
}
 
\maketitle
  
\begin{abstract}
 
Direct CP violation beyond the standard model can be produced in charged D 
decays to final states with a $K_S$ by small new physics contributions to 
the transitions $D^+ \rightarrow K^o X^+$, where $X^+$ denotes any positively 
charged hadronic state or transitions $D^+ \rightarrow K^o \bar K^o K^{(*)+}$, 
where $K^{(*)+}$ denotes any positive strange state. 
These transitions are doubly-Cabibbo suppressed and 
color suppresed in the standard model and branching ratios are experimentally 
observed to be suppressed by two orders of magnitude relative to the allowed 
$D^+ \rightarrow \bar K^o X^+$ or $D^+ \rightarrow \bar K^o \bar K^o K^{(*)+}$, 
branching ratio. An even smaller new physics contribution might produce an 
observable CP asymmetry in 
$D^\pm \rightarrow K_S X^\pm$ or $D^{\pm} \rightarrow K_S K_S K^{(*)\pm}$
decays. Since such asymmetries are easily checked in the early stages of any 
charm production experiment, it seems worth while to
check them before the opportunity is lost in later stages of the analysis, 
even if no theoretical model predicts such an asymmetry.

\end{abstract}

} 
 
\renewcommand{\baselinestretch}{1.2}
\setlength{\baselineskip}{16pt}
\vspace{2ex}

Since the Standard Model predictions for CP violation in charm decays 
are very small, the charm sector has been cited as a good place to test
the SM and to look for evidence for physics beyond the SM\cite{AITALA},
Some negative results were recently reported in a search for CP violation in 
certain singly Cabibbo suppressed decays. 
 
The same experiment might have been used to look for CP violation in the 
interference between Cabibbo favored and doubly-Cabibbo suppressed amplitudes.
This has been suggested by Bigi and Yamamoto\cite{Bigyam} and the basic physics
is discussed in detail in their paper.
Although these lead to final states with different strangeness, interference
between them can be observed in final states containing a $K_S$, as in
$D^\pm \rightarrow K_S X^\pm$ decays, where $X^\pm$ denotes any charged 
hadronic state. 
$$ D^\pm \rightarrow K^o X^\pm \rightarrow K_S X^\pm
 \eqno (1a)  $$
$$ D^\pm \rightarrow \bar K^o X^\pm \rightarrow K_S X^\pm
 \eqno (1b)  $$
The dominant standard model decay diagram is the Cabibbo favored
and color favored tree diagram shown in figure 1. The standard model also has
the doubly-Cabibbo suppressed and color suppresed tree diagram shown in figure
2 and the doubly-Cabibbo suppressed annihilation diagram shown in figure 3.
One can wonder about new-physics diagrams like those shown in figure 4 and 
figure 5 which respectively resemble the SM Cabibbo-suppressed tree and 
annihilation diagrams of figures 1 and 2 but go via some new physics boson or 
more complicated diagram instead of the $W$. Such a diagram might have a 
different weak phase from the dominant tree diagram of figure 1 and produce a 
direct CP violation in the decay modes $D^\pm \rightarrow K_S X^\pm$ via 
interference between the two diagrams. Such interference would produce an 
opposite CP violation in the decay modes $D^\pm \rightarrow K_L X^\pm$ and 
produce no violation in the total widths as required by CPT\cite{PENGRHO}. 

These final states had not previously been 
considered in the early selection stages of the data selection process in 
experiments like the Fermilab charm experiment E791 whose results were cited
above, and where the following observations have been made\cite{APPEL}: 

``One can imagine reaching a sensitivity of order $10^{-3}$ 
in comparing $D^\pm \rightarrow K_S 3 \pi$. That mode looks very 
interesting.  It is especially nice in being a self tagging mode (by charge) 
and having normalization signals which are very well known ($K \pi \pi$).
However, returning to these early selection stages at a later time 
has been estimated as requiring a large number of tapes 
to be mounted (say, 2000).  This would be a considerable effort.
Such an effort is hard to sell if there are no concrete predictions from
specific models for new physics." 

It is therefore of interest both for theorists
proposing new models to check whether they can provide such predictions, and for
experimenters planning new experiments to be aware of these possibilities at
early stages when the measurements are cheap and easy, even though no 
theoretical motivation exists at the time. 

Similarly, one can also consider decays into final states with two $K_S$; e.g.
the doubly-Cabibbo suppressed decays illustrated in figure 6,
$$ D^\pm \rightarrow K^o \bar K^o K^{(*)\pm} \rightarrow K_S K_S K^{(*)\pm}
 \eqno (2a)  $$
and the Cabibbo allowed decays
$$ D^+ \rightarrow \bar K^o \bar K^o K^{(*)+} \rightarrow K_S K_S K^{(*)+}
 \eqno (2b)  $$
$$ D^- \rightarrow K^o K^o K^{(*)-} \rightarrow K_S K_S K^{(*)-}
 \eqno (2c)  $$
where $K^{(*)+}$ denotes any positive strange state. 
Although there is no present model for new physics suggesting such  diagrams
a number of arguments support a search for this direct CP violation.

     1. Branching ratios are high\cite{PDG}; namely 6\% for $\bar K^o \rho^+$, 
8\% for $\bar K^o a_1^+$ and 59\% for $K^o + \bar K^o$ inclusive. Even with the 
loss of a factor of three in detecting $K^o$ via the decay chain 
$K^o \rightarrow K_s \rightarrow \pi^+ \pi^-$ and further experimental losses,
the remaining signal can be sufficiently large to see a signal. Even if 
nothing is found a sensible upper limit might be obtained that can shoot down 
some future theories. 

     2. The interference term is linear in the new physics and doubly-Cabibbo 
suppressed amplitude, while doubly-Cabibbo suppressed branching ratios are 
quadratic. The ratio of the experimental branching fractions for the analogous 
forbidden and allowed decays $D^+ \rightarrow K^+ \pi^+ \pi^-$ and 
$D^+ \rightarrow K^- \pi^+ \pi^+$ is given \cite{AITALADCF} as 
$(7.7 \pm 1.7 \pm 0.8)  \times 10^{-3 } \approx (3.0 \pm 0.8) \tan^4 \theta_C $.
This suggests that the 
forbidden decay is suppressed by one order of magnitude in amplitude and two
orders of magnitude in branching ratio. A new physics amplitude which is
two orders of magnitude below the dominant allowed amplitude and one order of
magnitude below the forbidden amplitude could give an interference contribution
to $D^\pm \rightarrow K_S X^\pm$ of several per cent and might produce an 
observable direct CP violation in the one per cent ball park. 

     3. The Cabibbo-favored amplitude leads to an exotic final state whereas
the doubly-suppressed and new-physics amplitudes lead to a non-exotic final 
state and can be enhanced by the presence of meson resonances\cite{CLOLIP12}.

     4. The presence of meson hadronic resonances near the $D$ mass has been 
pointed out and the possibility that they might influence charmed meson decays 
has been discussed. The nature of such resonances is still under
investigation and the possibility that they might be hybrid 
(quark-antiquark-gluon) states opens up new possibilities of enhanced 
contributions to penguin and annihilation diagrams which go via a $q \bar q G$
intermediate state\cite{CLOLIP12}. 

     5. CPT invariance requires that a final state must be a linear combination
of at least two eigenstates of the strong-interaction S-matrix with different
strong and weak phases in order for the observation of direct CP charge 
asymmetry\cite{PENGRHO}. This normally requires some nontrivial strong 
interaction rescattering. The present case is different because the two strong 
eigenstates have different strangeness and are not coupled by strong interaction
scattering. Interference is achieved by the use of weak interactions in a
detector that mixes $K^o$ and $\bar K^o$. The strong phases are expected to be
very different since one state is exotic and has no resonance phase, while the
other is non-exotic and is in the center of the resonance region. 

     6. CPT requires every CP charge asymmetry to be compensated by an opposite 
CP asymmetry elsewhere to give the same total widths. Here this compensation is 
automatic because all decays occur in pairs with $K_L$ in one mode and $K_S$ in 
the other or with $K_LK_L$ in one mode and $K_SK_S$ in the other. Any CP 
violating asymmetry in a $K_S$ mode is reversed in the $K_L$ mode, so that CPT 
restrictions are automatically satisfied without requiring complicated final 
state rescattering.

     7. The search is cheap. One only needs to separate the two charge states
that are observed anyway and give a result for the difference.
 
     We now describe these effects explicitly.
Let $A_f$, $A_{cs}$ and $A_{np}$ denote respectively the magnitudes of the 
Cabibbo-favored, the doubly-Cabibbo suppressed and new
physics contributions to the amplitude for a given decay $D^\pm \rightarrow 
K_S X^\pm$. Let $S$ and $W$ denote respectively the strong and weak phases for the 
dominant Cabibbo-favored amplitude for the case of 
the $D^+ \rightarrow K_S X^+$ decay, $S_{cs}$ denote the $difference$ between 
the strong phases of the Cabibbo suppressed and 
Cabibbo allowed amplitudes, and $S_{np}$ and $W_{np}$ denote respectively the 
$differences$ between the strong and weak phases of
the new physics and the strong and weak phases of
the Cabibbo allowed amplitudes, The total amplitude denoted by $ A^{\pm}$ for 
a given decay $D^\pm \rightarrow K_S X^\pm$ is then
$$ A^{\pm} = e^{i(S \pm W)}\cdot [A_f + e^{iS_{cs}}\cdot A_{cs} + e^{i(S_{np} 
\pm W_{np})}\cdot A_{np}] \eqno (3a)  $$
Thus
$$ |A^{\pm}|^2 \approx  A_f^2   + 2 A_f A_{cs} \cdot \cos S_{cs} + 2 A_f 
A_{np} \cdot \cos (S_{np} \pm W_{np}) \eqno (3b)  $$
and
$$ |A^-|^2 - |A^+|^2 \approx  4 A_f A_{np} \cdot \sin S_{np} \cdot \sin W_{np} 
\eqno (4)  $$

The number of events counted in a given experiment can then be written 
$$ N^{\pm} = |C^{\pm}|^2 \cdot |A^{\pm}|^2|  \approx |C^{\pm}|^2\cdot [A_f^2 + 
2 A_f A_{cs} \cos S_{cs} + 2 A_f A_{np} \cos (S_{np} \pm {W_{np}})] 
\eqno (5a)  $$
Where the normalization factors $C^{\pm}$ depend upon the conditions of the 
experiment, running time, acceptances, efficiencies, etc.

The statistical error is then given by
$$\delta N^{\pm} = \sqrt {N^{\pm}}  \approx C^{\pm}  A_f \eqno (5b)  $$

The observed $CP$ asymmetry is given by eq. (4), with a statistical error 
given by eq. (5b).
$$(A_{sym})_{CP} \equiv {{N^-}\over{|C^-|^2}}  
- {{N^+}\over{|C^+|^2}}  \approx 4 A_fA_{np}\cdot  \sin S_{np} \sin W_{np} 
\eqno (6a)  $$
$$\delta (A_{sym})_{CP} \approx  A_f \cdot 
{{\sqrt {|C^-|^{2} + |C^+|^{2}}}\over{C^+C^-}} \leq 
     A_f \cdot \sqrt {{{2} \over {C^+C^-}}}
\eqno (6b)  $$
The ratio of the CP signal to the statistical error is then 
$${{(Signal)_{CP}}\over{Statistical ~ Error}} =
{{(A_{sym})_{CP} }\over{\delta (A_{sym})_{CP}}} \approx
{{4 C^+C^-}\over{\sqrt {|C^-|^{2} + |C^+|^{2}}}} \cdot  
A_{np}\cdot  \sin S_{np} \sin W_{np} \eqno (7a)  $$
Substituting the inequality (6b) then gives
$${{(Signal)_{CP}}\over{Statistical ~ Error}} =
{{(A_{sym})_{CP} }\over{\delta (A_{sym})_{CP}}} \geq 
2\cdot \sqrt {2 C^+C^-} \cdot  
 A_{np}\cdot  \sin S_{np} \sin W_{np} \eqno (7b)  $$
The inequalities in eqs. (6b) and (7b) become equalities in cases where 
$C^+ \approx C^- \equiv C$ as in $D^+$ and $D^-$ decays 
produced in $e^+ e^-$ experiments. In this approximation  we can write
$$ {{(Signal)_{CP}}\over{Total ~ Rate}} =
{{C^2\cdot (A_{sym})_{CP} }\over{N}} \approx 4 \cdot
\left( {{A_{np}}\over{A_f}}\right )\cdot  
\sin S_{np} \sin W_{np} \eqno (8)  $$

Note that the ratio of the CP violation to the statistical 
error is independent of the dominant $A_f$ amplitude and depends only on the
new-physics amplitude $A_{np}$, even though the relative strength of the $CP$ 
violation is inversely proportional to $A_f$.
This illustrates a general feature of searches for $CP$ violation in a CP 
violating phase which occurs in an interference term between a dominant 
$CP$ conserving term and a small term. One might think that the small term would
be more important if its ratio to the dominant term is large. But this is not 
the case. The statistical significance of an observed violation is independent 
of the relative strengths of the dominant term and the small term. 

In cases where $C^+ \not= C^- $ it is necessary to obtain the ratio of these
normalization factors from another decay mode which will presumably have better
statoistics and a statistical error negligible in comparison with those of the 
decays under investigation. 

It is interesting to compare the doubly-Cabibbo suppressed decays into modes
with neutral kaons and the corresponding decays into modes with charged kaons. 
The branching ratios into the charged modes have no interference with allowed
decays and therefore give a measure of the doubly-Cabibbo suppressed amplitudes.
However, other factors must be taken into account in order to interpret these 
results.

For the quasi-two-body decays with a single kaon (1), the corresponding 
decays with a charged kaon are
$$ D^\pm \rightarrow K^\pm X^o  \eqno (9)  $$
These decays are color favored, in contrast with the doubly-Cabibbo-suppressed
decays (1) into modes with neutral kaons which are also color suppressed.

For the three-kaon decays (2a) with two neutral kaons, the corresponding 
decays with a charged kaon pair are
$$ D^\pm \rightarrow K^+ K^- K^{(*)\pm}  \eqno (10)  $$
In contrast to the decay modes with neutral kaon pairs (2a) which can be 
produced by the dominant doubly-Cabibbo-suppressed and color-favored tree 
diagram shown in figure 6, the charged decay modes (10) cannot be produced by 
this diagram, which contains a $d \bar d$ pair instead of the required 
$u \bar u$ pair. The charged transition (10) can only be produced by an 
annihilation diagram or by an additional final state strong charge-exchange 
scattering; i.e. a $d \bar d \rightarrow u \bar u$ transition following the 
tree diagram of figure 6. 

The search for CP violation in interference between Cabibbo-favored and
Doubly-Cabibbo-suppressed decays has been suggested by Gronau, Wyler, Dunietz 
and others\cite{GRONAU} for $B \rightarrow$ charm decays where CP effects are 
suggested by the standard model. In the charm decays discussed here there is no 
prediction from the standard model for direct CP violation, but branching 
ratios are much higher and there is the possibility of additional effects due 
to resonances which are absent at the B mass.

\acknowledgments
It is a pleasure to thank Jeffrey Appel, Edmond Berger, Karl Berkelman, John
Cumalat, Yuval Grossman, Yosef Nir and J. G. Smith for helpful discussions and 
comments.
 This work was partially supported by the German-Israeli Foundation
for Scientific Research and Development (GIF).
 
{
\tighten

}

{\begin{figure}[htb]
$$\beginpicture
\setcoordinatesystem units <\tdim,\tdim>
\stpltsmbl
\putrule from -25 -30 to 50 -30
\putrule from -25 -30 to -25 30
\putrule from -25 30 to 50 30
\putrule from 50 -30 to 50 30
\plot -25 -20 -50 -20 /
\plot -25 20 -50 20 /
\plot 50 0 120 -20 /
\plot 50 -20 120 -40 /
\photonru 50 20 *3 /
\plot 120 40 90 20 120 20 /
\put {$c$} [b] at -50 25
\put {$\overline{d}$} [t] at -50 -25
\put {$u$} [l] at 125 40
\put {$\overline{d}$} [l] at 125 20
\put {$s$} [l] at 125 -20
\put {$\overline{d}$} [l] at 125 -40
\put {$\Biggr\}$ $X^+$} [l] at 135 30
\put {$\Biggr\}$ {$\overline{K}^o  \rightarrow K_S$} } [l] at 135 -30
\put {$W$} [t] at 70 15
\setshadegrid span <1.5\unitlength>
\hshade -30 -25 50 30 -25 50 /
\linethickness=0pt
\putrule from 0 0 to 0 60
\endpicture$$
\caption{\label{fig-1}} \hfill Standard Model Cabibbo and Color favored diagram.
 \hfill~ \end{figure}}
 
{\begin{figure}[htb]
$$\beginpicture
\setcoordinatesystem units <\tdim,\tdim>
\stpltsmbl
\putrule from -25 -30 to 50 -30
\putrule from -25 -30 to -25 30
\putrule from -25 30 to 50 30
\putrule from 50 -30 to 50 30
\plot -25 -20 -50 -20 /
\plot -25 20 -50 20 /
\plot 50 20 120 40 /
\plot 50 -20 120 -40 /
\photonru 50 0 *3 /
\plot 120 20 90 0 120 -20 /
\put {$c$} [b] at -50 25
\put {$\overline{d}$} [t] at -50 -25
\put {$d$} [l] at 125 40
\put {$\overline{s}$} [l] at 125 20
\put {$u$} [l] at 125 -20
\put {$\overline{d}$} [l] at 125 -40
\put {$\Biggr\}$ $K^o \rightarrow K_S$} [l] at 135 30
\put {$\Biggr\}$  $X^+$} [l] at 135 -30
\put {$W$} [t] at 70 -5
\setshadegrid span <1.5\unitlength>
\hshade -30 -25 50 30 -25 50 /
\linethickness=0pt
\putrule from 0 0 to 0 60
\endpicture$$
\caption{\label{fig-2}} \hfill SM Color and Double-Cabibbo suppressed diagram.
\hfill~ \end{figure}}

{\begin{figure}[htb]
$$\beginpicture
\setcoordinatesystem units <\tdim,\tdim>
\stpltsmbl
\putrule from -25 -30 to 50 -30
\putrule from -25 -30 to -25 30
\putrule from -25 30 to 50 30
\putrule from 50 -30 to 50 30
\plot -25 -20 -50 -20 /
\plot -25 20 -50 20 /
\plot -25 -20  -15 0 -25 20   /
\plot 50 20 120 40 /
\plot 50 -20 120 -40 /
\plot 50 20 25 0 50 -20  /
\photonru -15 0 *3 /
\springru 50 0 *3 /
\plot 120 20 90 0 120 -20 /
\put {$c$} [b] at -50 25
\put {$\overline{d}$} [t] at -50 -25
\put {$u$} [l] at 125 40
\put {$\overline{d}$} [l] at 125 20
\put {$d$} [l] at 125 -20
\put {$\overline{s}$} [l] at 125 -40
\put {$\Biggr\}$  $X^+$}  [l] at 135 30
\put {$\Biggr\}$ $  K^o \rightarrow K_S$} [l] at 135 -30
\put {$G$} [t] at 70 -5
\put {$W$} [t] at 0 -5
\setshadegrid span <1.5\unitlength>
\hshade -30 -25 50 30 -25 50 /
\linethickness=0pt
\putrule from 0 0 to 0 60
\endpicture$$
\caption{\label{fig-3}} \hfill SM Annihilation. $G$ denotes any number of
gluons. \hfill~ \end{figure}}

{\begin{figure}[htb]
$$\beginpicture
\setcoordinatesystem units <\tdim,\tdim>
\stpltsmbl
\putrule from -25 -30 to 50 -30
\putrule from -25 -30 to -25 30
\putrule from -25 30 to 50 30
\putrule from 50 -30 to 50 30
\plot -25 -20 -50 -20 /
\plot -25 20 -50 20 /
\plot 50 20 120 40 /
\plot 50 -20 120 -40 /
\photonru 50 0 *3 /
\plot 120 20 90 0 120 -20 /
\put {$c$} [b] at -50 25
\put {$\overline{d}$} [t] at -50 -25
\put {$d$} [l] at 125 40
\put {$\overline{s}$} [l] at 125 20
\put {$u$} [l] at 125 -20
\put {$\overline{d}$} [l] at 125 -40
\put {$\Biggr\}$ $K^o \rightarrow K_S$} [l] at 135 30
\put {$\Biggr\}$  $X^+$} [l] at 135 -30
\put {$???$} [t] at 70 -5
\setshadegrid span <1.5\unitlength>
\hshade -30 -25 50 30 -25 50 /
\linethickness=0pt
\putrule from 0 0 to 0 60
\endpicture$$
\caption{\label{fig-4}} \hfill New Physics tree  diagram.
\hfill~ \end{figure}}

{\begin{figure}[htb]
$$\beginpicture
\setcoordinatesystem units <\tdim,\tdim>
\stpltsmbl
\putrule from -25 -30 to 50 -30
\putrule from -25 -30 to -25 30
\putrule from -25 30 to 50 30
\putrule from 50 -30 to 50 30
\plot -25 -20 -50 -20 /
\plot -25 20 -50 20 /
\plot -25 -20  -15 0 -25 20   /
\plot 50 20 120 40 /
\plot 50 -20 120 -40 /
\plot 50 20 25 0 50 -20  /
\photonru -15 0 *3 /
\springru 50 0 *3 /
\plot 120 20 90 0 120 -20 /
\put {$c$} [b] at -50 25
\put {$\overline{d}$} [t] at -50 -25
\put {$u$} [l] at 125 40
\put {$\overline{d}$} [l] at 125 20
\put {$d$} [l] at 125 -20
\put {$\overline{s}$} [l] at 125 -40
\put {$\Biggr\}$  $X^+$}  [l] at 135 30
\put {$\Biggr\}$ $  K^o \rightarrow K_S$} [l] at 135 -30
\put {$G$} [t] at 70 -5
\put {$???$} [t] at 0 -5
\setshadegrid span <1.5\unitlength>
\hshade -30 -25 50 30 -25 50 /
\linethickness=0pt
\putrule from 0 0 to 0 60
\endpicture$$
\caption{\label{fig-5}} \hfill New Physics Annihilation. 
 \hfill~ \end{figure}}

{\begin{figure}[htb]
$$\beginpicture
\setcoordinatesystem units <\tdim,\tdim>
\stpltsmbl
\putrule from -25 -30 to 50 -30
\putrule from -25 -30 to -25 30
\putrule from -25 30 to 50 30
\putrule from 50 -30 to 50 30
\plot -25 -20 -50 -20 /
\plot -25 20 -50 20 /
\plot 50 10 120 0 /
\plot 50 -20 120 -40 /
\photonru 50 20 *3 /
\plot 120 40 90 20 120 30 /
\springru 50 -15 *3 /
\plot 120 -10 90 -15 120 -30 /
\put {$c$} [b] at -50 25
\put {$\overline{d}$} [t] at -50 -25
\put {$u$} [l] at 125 40
\put {$\overline{s}$} [l] at 125 30
\put {$d$} [l] at 125 10
\put {$s$} [l] at 125 -30
\put {$\overline{s}$} [l] at 125 -10
\put {$\overline{d}$} [l] at 125 -40
\put {$\Biggr\}$ $K^+$} [l] at 135 35
\put {$\Biggr\}$ $K^o \rightarrow K_S$} [l] at 135 0
\put {$\Biggr\}$ {$\overline{K}^o  \rightarrow K_S$} } [l] at 135 -35
\put {$W$} [t] at 70 30
\setshadegrid span <1.5\unitlength>
\hshade -30 -25 50 30 -25 50 /
\linethickness=0pt
\putrule from 0 0 to 0 60
\endpicture$$
\caption{\label{fig-6}} \hfill Standard Model Double-Cabibbo Suppressed
$D \rightarrow 3K $ diagram.
 \hfill~ \end{figure}}

\end{document}